\documentclass[prd,twocolumn,superscriptaddress,showpacks]{revtex4}
\usepackage{empheq}
\usepackage{amsmath}
\usepackage{amssymb}
\usepackage{dsfont}
\usepackage{graphicx}
\usepackage{hyperref}
\usepackage{stackrel}
\usepackage{color}
\usepackage{comment}

\def \ee{\end{equation}}
\def \be{\begin{equation}}
\def \eea{\end{eqnarray}}
\def \bea{\begin{eqnarray}}

\begin{document}

\title{Path integral of the relativistic point particle
in Minkowski space}

\author{Benjamin Koch}
\email{benjamin.koch@tuwien.ac.at}
\affiliation{Institut f\"ur Theoretische Physik,
 Technische Universit\"at Wien,
 Wiedner Hauptstrasse 8-10,
 A-1040 Vienna, Austria.}
\author{Enrique Mu\~noz}
\affiliation{Pontificia Universidad Cat\'olica de Chile \\ Instituto de F\'isica, Pontificia Universidad Cat\'olica de Chile, \\
Casilla 306, Santiago, Chile}
\email{munozt@fis.puc.cl}
\begin{abstract}
In this article, we analyze the fundamental global and local symmetries involved
in the action for the free relativistic point particle in Minkowski space. 
Moreover, we identify a hidden
local symmetry, whose explicit consideration and factorization utilizing of a Fujikawa prescription, leads to the construction of relativistic
two point correlation functions that satisfy the Chapman-Kolmogorov identity. 
By means of a detailed
topological analysis, we find three different relativistic correlation functions (orthochronous, space-like, and Feynman)
which are obtained from the exclusive integration of paths within different sectors in Minkowski space.
Finally, the connection of this approach to the Feynman checkerboard construction is explored.
\end{abstract}

\maketitle

\section{Introduction}

\subsection{The context}

Symmetries are key ingredients of our construction recipes for fundamental field theories.
Local symmetries (gauge symmetries) give us a method to study interactions while global symmetries 
give us a way to describe conserved charges via the Noether theorem.
For this article, two cases will be particularly important.
The first case is global Lorentz invariance which leads
to conserved quantities such as mass, momentum, angular momentum, and center-of-mass energy.
This symmetry is an essential part of all fundamental field theories and it
is intimately linked to the spin-statistics relation of fundamental particles.
The second case is a gauge symmetry known as diffeomorphism invariance,
which is, for example, the underlying symmetry of the theory of general relativity (GR).

Another indispensable ingredient of any fundamental physical theory is
that it needs to be quantized. While the combination of global Lorentz
invariance with the principles of quantum mechanics works out nicely, 
a consistent quantization of general relativity is still an open problem.
It is widely believed that this problem is somehow linked to diffeomorphism
invariance as the underlying gauge symmetry of GR. 
Unfortunately, gravity in (1+3) dimensions is a very complicated theory with this symmetry.
Thus, the community has put a large effort into studying and understanding
somewhat simpler theoretical realizations of this symmetry, such as for example
Euclidean quantum gravity \cite{tHooft:1974toh,Lauscher:2001ya,Reuter:2001ag,Litim:2003vp},
or quantum gravity with Lorentz signature, but in lower dimensional systems~\cite{tHooft:1988qqn,Ashtekar:1989qd,Carlip:1998uc}.
The most simple theoretical system with diffeomorphism invariance is 
the action of the relativistic point particle.  Attempts to
quantize the straight action of the relativistic point particle (\ref{actionRPP})
lead to substantial difficulties~\cite{Teitelboim:1982,Henneaux:1982ma,Redmount:1990mi,Fradkin:1991ci,Kleinert:book,Polchinski:book,Padmanabhan:1994}, 
such as the loss of the Champman-Kolmogorov identity.
In the literature, these problems are circumvented 
by using a different, but classically equivalent, action~\cite{Brink:1976,Brink:1977,Fradkin:1991ci},
by introducing a modified theory of probabilities~\cite{Jizba:2008,Jizba:2010pi,Jizba:2011wg}, by 
the restriction to particular manifolds~\cite{Fukutaka:1986ps}, by the use of approximations~\cite{Padmanabhan:1994},
or by a description based on a subtle limit of a quantum field theoretical prescription~\cite{Padmanabhan:2017bll}.

In a series of previous papers, we have shown that 
these problems with the quantization of the action (\ref{actionRPP})
can actually be solved by taking into consideration
yet another (hidden) local symmetry, corresponding the
invariance of the square of the 4-velocity $(d{x}^{\mu}/d\lambda) (d{x}_{\mu}/d\lambda)$ under Lorentz boosts and rotations. Despite this is not a gauge symmetry, 
we have shown from three different and complementary approaches that
it can nevertheless be factored out in the path-integral (PI) construction, thus restoring the Chapman-Kolmogorov property in the resulting Two Point Correlation Functions (TPCFs).
We developed this new method in three complementary PI approaches:
\begin{itemize}
\item
A PI over the Hamiltonian action and the corresponding constraint analysis~\cite{Koch:2019vxw}
\item
A direct PI in Euclidean space~ \cite{Koch:2017nha}
\item 
A formal functional path integral~\cite{Koch:2017bvv}
\end{itemize}
These results were either obtained in Euclidean space or
with abstract formal methods, but a detailed analysis of
the much richer structure of Minkowski space is missing.

This article closes this gap by discussing the 
path integral quantization of the RPP in Minkowski space-time,
while paying explicit attention to the causal structure of virtual paths.
By doing so, new insights can be obtained
on how different types of paths lead to different TPCFs.
Such a discussion is particularly interesting since it gives
evidence that also in quantum gravity the path integral
quantization will give results that are sensitive to the
causal structure of the system.

This article is organized as follows.
In the next two subsections of the introduction, we will 
introduce the relativistic point particle action and the conceptual
steps which have to be considered in the quantization of this system.
The notation of this introduction closely follows Ref.~\cite{Koch:2017nha}.
In the next section, the quantization procedure will be realized.
First, the discrete action will be defined, then the two step TPCF will be calculated
before this result is generalized to the $N+1$ step TPCF.
This discussion will be done by summing exclusively over causal time-like (orthochronous) paths.
A generalization with a comparison to space-like and time-like non-causal paths follows 
in the next subsection.
Then, a particular case of this calculation is considered, where 
a spatial flip symmetry is removed from the construction.
This results in the famous Feynman checkerboard.
After generalizing the TPCFs with flip symmetry to $1+d$ dimensions
and commenting on the higher dimensional checkerboard
we summarize our findings in the conclusions.

\subsection{The relativistic point particle}

The action for a relativistic point particle in Minkowski space
with the metric signature $g^{00}=+1$, $g^{ii} =-1$ for $i = 1,2,3$ is
\be\label{actionRPP}
S=-\int_{\lambda_i}^{\lambda_f} d\lambda \cdot m \sqrt{\left(\frac{d  x^\mu}{d\lambda} \frac{d  x_\mu}{d\lambda}\right)},
\ee
where $x^\mu(\lambda_i)=x^\mu_i$ and $ x^\mu(\lambda_f)= x^\mu_f$.
This is simply the mass $(m)$ times the geometric length of a given path $\mathcal{P}$
in Euclidean space or, equivalently, the mass times the total interval between two events in Minkowski space.
This action and its corresponding Lagrangian are equipped with several symmetries
which will be important for the formulation of a consistent path integral \cite{Kleinert:book,Polchinski:book}.
\begin{itemize}
\item[({\bf a})] Global Poincar\'e invariance: This can be seen from the fact that the action
is invariant under global rotations, boosts, and shifts of the coordinate system.
\item[({\bf b})] Local Lorentz invariance: This means that the Lagrangian is invariant under
local rotations and boosts of the vector $(d x^\mu)/(d\lambda)$ at any point along the trajectory. Since 
this symmetry vanishes at the classical level and thus
its role on the construction of
the TPCFs is typically neglected in the literature, we termed it a ``hidden symmetr''.
We presented a formal argument on why this symmetry, which is not a classical gauge symmetry, is important 
in this given context in \cite{Koch:2019vxw,Koch:2017nha,Koch:2017bvv}. 
In \cite{Koch:2019vxw}, we further showed in the Hamiltonian action formulation that there is a non-trivial constraint
associated with this symmetry. 
\item[({\bf c})] Weyl invariance: This means that the Lagrangian does not depend on the way that
$\lambda$ parametrizes a path $\mathcal{P}$. The change to any other monotonous function $\tilde \lambda(\lambda)$ would 
leave the Lagrangian invariant.
\end{itemize}
In the following the symmetry ({\bf{a}}) will be used to choose the coordinate system such
that $x^{\mu}_i=0$ and that $x^{\mu}_f$ is different from zero in only one component.
The symmetries ({\bf{b}}) and ({\bf{c}}) are symmetries which have to be
treated with care when it comes to realizing an integral over different paths since
two seemingly different paths could be actually physically equivalent. 
The over-counting of physically equivalent paths
would result in a wrong weight of some paths with respect to others.

\subsection{General considerations on the explicit form of the PI measure}

A naive, straightforward definition of a path integral
for the relativistic point particle fails to satisfy the Kolmogorov condition for transition probability amplitudes. As shown in \cite{Koch:2019vxw,Koch:2017nha,Koch:2017bvv},
this is due to the overcounting arising from the symmetries {\bf{b)}}, and {\bf{c)}}, that needs to be properly factored out either by geometric considerations \cite{Koch:2019vxw,Koch:2017nha}
or by a group theoretical analysis involving the Fadeev-Popov method \cite{Koch:2017bvv}.

\begin{figure}[hbt]
   \centering
\includegraphics[width=0.4\textwidth]{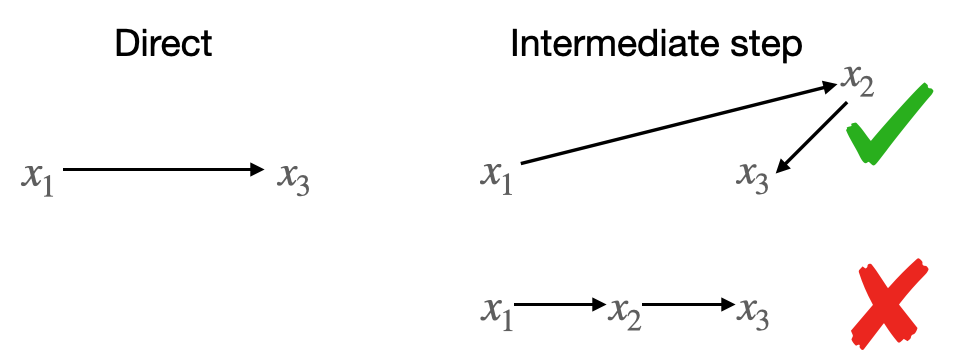}
  \caption{\label{OCfig} Exemplification of possible overcounting of one path
  with two intermediate steps, which is already counted in the PI with one intermediate step. 
  }
\end{figure}

Following the geometric approach, let us consider as an example
the case of two intermediate points $n=2$.
All configurations where the position $ x^\mu_1$ is on the classical
path between $x^{\mu}_i\rightarrow x^{\mu}_2$ correspond actually to the same path
$x^{\mu}_i\rightarrow x^{\mu}_2\rightarrow x^{\mu}_f$ as it is shown in figure \ref{OCfig}.
%
Strictly speaking, they should only be counted once, but according to 
a naive counting this path would
be counted multiple times.
In many cases, this type of overcounting is not relevant, since the number of paths that
are not overcounted grows much faster with $n$ and $D$ than the number of paths where this overcounting occurs.
Thus, an improved definition of the path integral can be
given by
\bea\label{measure2}
&&K(x_i^\mu,x_f^\mu)\equiv \lim_{n\rightarrow \infty}\left. \sum_{j=1}^n K^{(j)}(x_i^\mu,x_f^\mu)\right|_{NOC}\\ \nonumber
&&\equiv\left.
\lim_{n\rightarrow \infty} \sum_{j=1}^n \left(\prod_{l=1}^j \int  dx_l^D\right|_{NOC}\right)\Xi_j
\exp\left[-i S_j\right] \\ \nonumber
&&=\left.\int dx^D_1\right|_{NOC}\Xi_1
\exp\left[-i S_1\right]\nonumber\\ 
&&+ \left.\int dx^D_1\int dx^D_2\right|_{NOC} \Xi_2
\exp\left[-i S_2\right] 
+\dots,\nonumber
\eea
where $|_{NOC}$ stands ``integrate and sum without overcountig'' in the sense of the symmetries 
described in {\bf{b)}} and {\bf{c)}}.
Further, $\Xi_i$ is the Fujikawa determinants assuring the invariance of the
measure under these symmetries~\cite{Fujikawa:1979ay}. 
The action for a path with $q$ intermediate steps (defining $x_{j=0}^{\mu} = x_i^{\mu}$, $x_{j=q+1}^{\mu} = x_f^{\mu}$) is given by the recursive relation
\be
S_q= -m \sum_{j=1}^{q+1}\sqrt{(x_{j}-x_{j-1})^\mu (x_{j}-x_{j-1})_\mu}.
\ee
Relation (\ref{measure2}) will be used for the calculation
of the path integral of the relativistic point particle in Minkowski space.
The following discussion will be done in $D=1+1$ dimensions
and a generalization to an arbitrary number of dimensions will be given in section \ref{sec:high_dim}.

\section{The TPCF}

We shall first consider a geometrical analysis of the problem, closely following Ref.\cite{Koch:2017nha}, starting with the calculation of the one-slice TPCF
and then generalizing to the $n$-slice TPCF.

\subsection{Action}

For the one-slice TPCF, based on the global Poincar\'e invariance (a), we can choose $x_i^{\mu}=(0,0)$, going to $x_f^{\mu}=(t_f,0)$
via $x_1^\mu=(t_1,x_1)$. Since the action 
of the relativistic point particle does not depend on the sign of $t_f$,
we will work with purely positive time differences $t_f=|t_f|$.
The action for this configuration is
\bea\label{S1}
 S&=&S_1=- m\left(\sqrt{(x_f -x_1)_\mu  (x_f- x_1)^\mu}\right.\nonumber\\
 &&\left.+ \sqrt{(x_1- x_i)_\mu  (x_1- x_i)^\mu}\right).
\eea
The minimum for (\ref{S1}) is obtained for the classical motion along a straight line
\be\label{S0}
S_{cl}=-m\sqrt{(x_f- x_i)_\mu  (x_f- x_i)^\mu}.
\ee
When considering only orthochronous paths which lie in the time-like
future light cone of $x_i$ and the time-like past light cone of
$x_f$, the maximum value for (\ref{S1})
is zero.
%
 \begin{figure}[hbt]
   \centering
\includegraphics[width=0.4\textwidth]{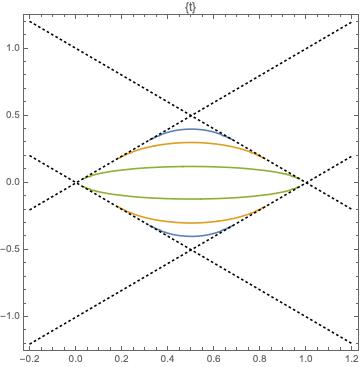}
\includegraphics[width=0.4\textwidth]{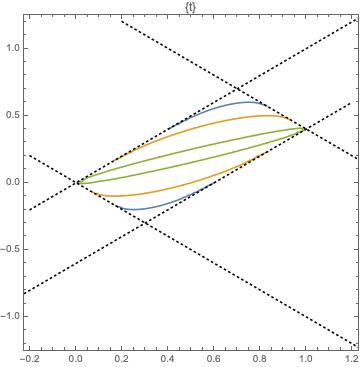}
  \caption{\label{LT0} 
 Contours of constant action (\ref{S1}). The light cones
 of the initial and final point are black dotted lines. 
 The upper figure is for $x_f=(1,0)$, the lower figure is for $x_f=(1,0.4)$. 
 For both figures $x_i(0,0)$ and $m=1$ was chosen. }
\end{figure}
In most of the following discussions,
we will expect the TPCFs to be Lorentz invariant. 
In those cases one can, without loss of generality,
choose a reference system where $x_i^{\mu}=(0,0)$
and $x_f^{\mu}=(t_f,0)$.

\subsection{The two step TPCF}
\label{subsecCalc}
With the action (\ref{S1}) one can define the
TPCF with one intermediate integration point
\be\label{K1}
K^{(1)}={\mathcal{N}}\cdot  \int  \int_V dt_1 dx_1 \Xi_1 e^{+i S_1},
\ee
where the integration volume $V$ is given by the overlap
of the future light cone of $x_i$ with the past light cone of $x_f$.
Note that for all ``reasonable'' paths the action (\ref{actionRPP}) is negative,
which is why the exponential in (\ref{K1}) was defined with a plus sign instead 
of a minus sign.
The term $\Xi_1$ in (\ref{K1}) is the Fujikawa measure for these integrals.
The role of $\Xi_1$ is
\begin{itemize}
\item[$\alpha$)] It has to render the path integral measure
invariant under changes in the normalized gauge parameter e.g. $\hat t_1$,
which runs from $0 \dots 1$.
This can be done by choosing the factor $\Xi(t_1)$ such that the product
$ \int_0^1 d\hat t_1 J \Xi_1$ is independent of $\hat t_1$.
Alternatively, this can be achieved by choosing a value of $\hat t_1$
where the Jacobian determinant of the integral is independent of changes in $\hat t_1$.
\item[$\beta$)] 
The above procedure $\alpha )$ leaves a freedom in terms of a multiplicative factor of the 
action value $S^b$.
It is imposed that $\Xi$ has to multiply the measure with the necessary power of $S^b$ in order
to keep the integrand no-zero and finite for $S\rightarrow 0$.
\item[$\gamma$)]
The above procedures $\alpha )$ and $\beta )$ leave a freedom in terms of a multiplicative factor
containing the external constants $m$ and $t_f$. Without loss of generality,
$m$ can be chosen to be one and the external power of $t_f^c$ is, like always, determined
from an additional normalization condition. 
\end{itemize}

As described above, a crucial step of
the recipe for the calculation of path integrals
with the hidden local symmetry is to rewrite the integral
in terms of the action as an integration variable.
For the TPCF (\ref{K1}) this can be achieved by
a coordinate change $\{x_1\rightarrow S, \;t_1 \rightarrow \hat t_1\}$ with
\bea\label{coordtrans}
x_1&=&\frac{\sqrt{(t_f^2-(S/m)^2)((S/m)^2-(t_f-2t_1)^2)}}{2 (S/m)}\\
t_1&=&\frac{(S/m)^2}{2 t_f}\hat t_1,
\eea
where the integration goes for $S:  -t_f m\dots 0$ and $\hat t_1: 0 \dots 1$.
However, for convenience, the minus sign of the action $S$ can  be absorbed in the redefinition of the
 integration variable $ \tilde S= -S$ with the corresponding change of the boundary 
 $- t_f m\rightarrow t_f m$.
The Jacobian determinant
for the applied transformations is
\bea\label{jac}
J&=&\frac{1}{\sqrt{(m t_f)^2-\tilde S^2}}\nonumber\\
&&\cdot 
\frac{\tilde S^4 (\hat t_1^2-1)-2 m^2\tilde S^2 \hat t_1 t_f^2 + m^4 t_f^4}{4 m^3 t_f \sqrt{\tilde S^2-((t_f m)-\tilde S^2\hat t_1 /(m t_f))^2}}.
\eea
Thus,
\be
K^{(1)}= {\mathcal{N}}\cdot 
\int^{t_f m}_0 d \tilde S 
\int_{0}^{1} d\hat t_1 
J \Xi_1 e^{-i \tilde S},
\ee
Since from condition $\alpha )$ one demands 
\be
\frac{d}{d\hat t_1}J\cdot \Xi_1=0,
\ee
then one can simply choose the Fujikawa determinant as the inverse of the $\hat t_1$ dependent
second term in Eq.~(\ref{jac}),
\be
\Xi_1=\frac{4 m^3 t_f \sqrt{\tilde S^2-((t_f m)-\tilde S^2\hat t_1 /(m t_f))^2}}{\tilde S^4 (\hat t_1^2-1)-2 m^2\tilde S^2 \hat t_1 t_f^2 + m^4 t_f^4}.
\ee
Equivalently one can also apply the 
Principle of Minimal Sensitivity (PMS)~\cite{Stevenson:1980du,Stevenson:1981vj,Stevenson:1982qw,Koch:2014joa}
by choosing the optimal
(symmetric and stable) value $\hat t_1^{opt}=\frac{m^2t_f^2}{\tilde S^2}$, as outlined in $\alpha )$. This alternative approach will be
used in higher spatial dimensions, as shown in section~\ref{sec:high_dim}.

A subsequent normalization with powers of $\tilde S$ and $t_f$ 
as mentioned in $\beta )$ and $\gamma )$ gives the TPCF
\begin{widetext}
\bea\label{res2}
K^{(1)}_{O}(0,t_f)={\mathcal{N}} \int_0^{t_f m} d\tilde S \frac{m}{\sqrt{(m t_f)^2-(\tilde S)^2}} \exp{(-i \tilde S)}
\left(\int_0^1 d\hat t_1\right)=
{\mathcal{N}} \frac{m \pi}{2}  \left(J_0(t_f m) - i\,H_0(t_f m)\right),
\eea
\end{widetext}
where $J_0$ is the Bessel function of the first kind and $H_0$ is the Struve function. This result can be stated in an explicitly relativistic invariant form,
by noticing that $m t_f = m \sqrt{\left( x_f - x_i\right)^{\mu}\left( x_f - x_i\right)_{\mu}} \equiv m |x_f - x_i|$, such that
\bea\label{oc2}
K^{(1)}_{O}(x_i^{\mu},x_f^{\mu})&=&{\mathcal{N}} \frac{m \pi}{2}  \left(J_0(m|x_f - x_i|)\right.\nonumber\\
&&\left.- i\,H_0(m |x_f - x_i|)\right).
\eea
The subindex ``$O$'' refers to time-like-orthochronous, since for this 
TPCF all virtual paths were time-like and respected the time ordering
of $t_f>t_1>0$. 

We notice that both functions $J_0(z)$ and $H_0(z)$ are independent solutions of the Bessel differential equation. Therefore, as proved in Appendix~\ref{AppA},
when $z = m|x|$ they are also independent solutions of the Klein-Gordon equation. One can thus conclude that the TPCF is itself a solution
of the Klein-Gordon equation for $|x|>0$, i.e.
\begin{eqnarray}
\left( \square_x + m^2 \right) K^{(1)}_{O}(m|x|) = 0.
\end{eqnarray}

In a later section, the relation between this TPCF and the Feynman propagator 
of a scalar field will be discussed in more detail. 

\subsection{The $N+1$ step TPCF}
\label{sec:nprop}

The full TPCF is defined as the sum of all possible TPCFs $K_O^{(n)}$
with $n$ intermediate steps
\be\label{Kfull}
K_O=\sum_{n=1}^{\infty} K_O^{(n)}|_{NOC}.
\ee
Since $K_O^{(1)}$ was calculated in the previous section,
let's now continue with $K_O^{(2)}$, by showing that
the contribution of this TPCF is zero due to the ``no overcounting''
condition $|_{NOC}$.

\begin{figure}[hbt]
 \begin{minipage}{5.5cm}
\centerline{\includegraphics[width=5.5cm]{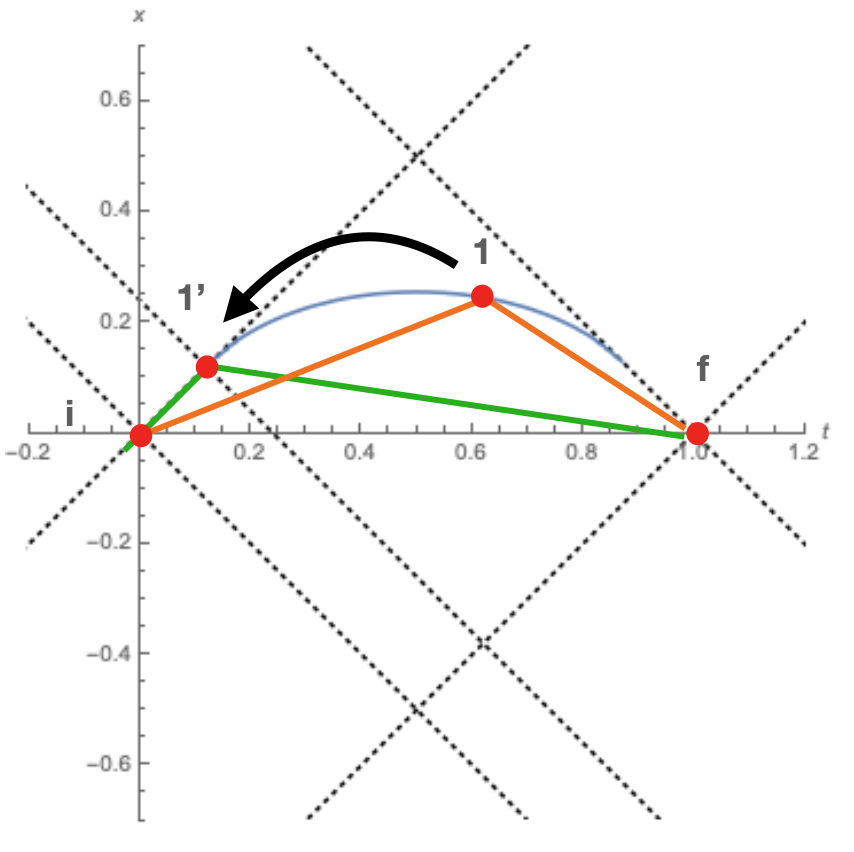}}
\end{minipage}
\ \
\hfill  \begin{minipage}{5.5cm}
\centerline{\includegraphics[width=5.5cm]{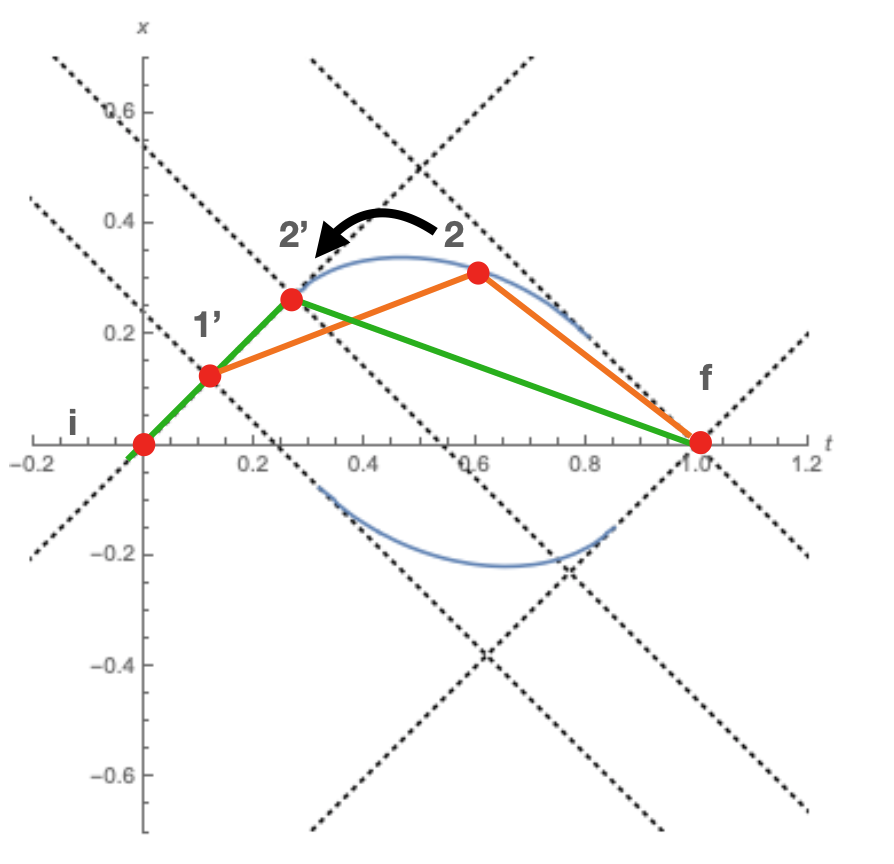}}
\end{minipage}
\ \
\hfill 
\begin{minipage}{5.5cm}
\centerline{\includegraphics[width=5.5cm]{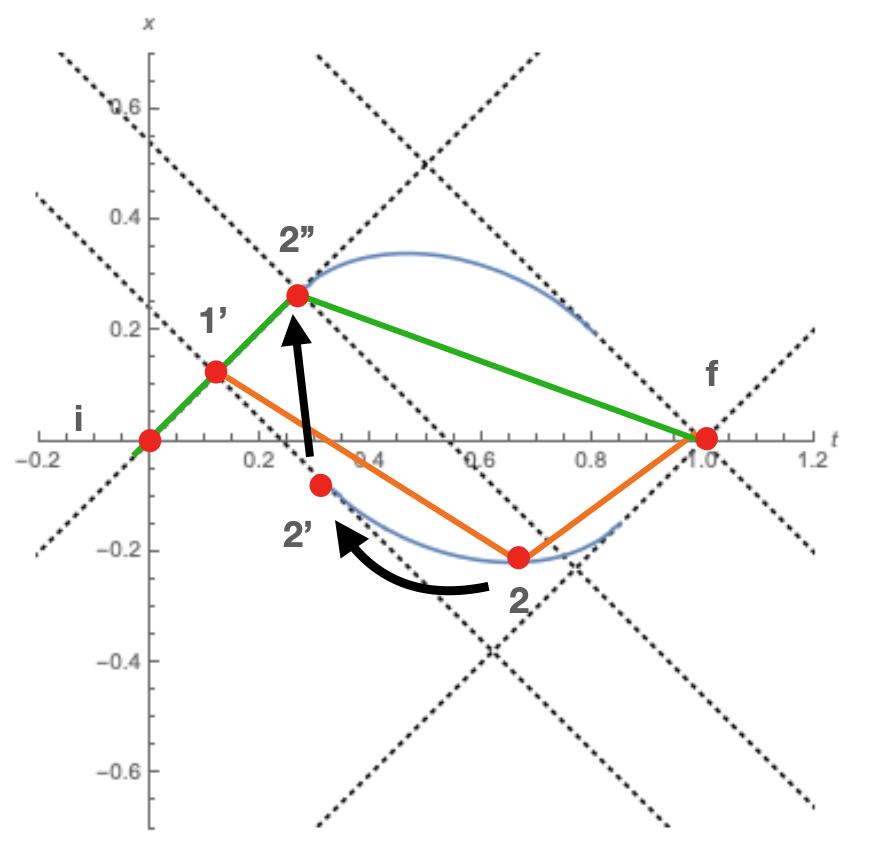}}
\end{minipage}
\ \
\hfill 
  \caption{\label{fig:twostep1}The graphical sequence of steps (see the main text) leading to the construction of the orthochronous TPCF. 
  }
\end{figure}

In this proof, a ``spatial flip'' symmetry will be assumed.
The $K_O^{(2)}$ TPCF can be constructed by the following steps:
\begin{itemize}
\item
First choosing an
arbitrary intermediate point $x_1^\mu$.  This situation is 
shown in the panel of figure \ref{fig:twostep1}. This point corresponds
to an action value $S_1 = S_{cl}(0,x_1^\mu)+S_{cl}(x_1^\mu, x_f^\mu)$.
However, due to the symmetry {\bf{b)}} from all the points with the same action value, indicated 
by the blue contour in the first panel of figure \ref{fig:twostep1},
only one has to be counted. As indicated in this panel we will
choose the point ${x'}^\mu_1$ which is light-like with the initial position $0$.
\item
 Having fixed the point ${x'}_1^\mu$, one chooses an arbitrary  second point
$x_2^\mu$, which is in the future light cone of ${x'}_1^\mu$ and in the past
light cone of $x_f^\mu$. This situation is shown in the second panel of 
figure \ref{fig:twostep1}. This choice corresponds to a different action, while now
keeping ${x'}_1^\mu$ fixed, again due to symmetry (b), the point  $x_2^\mu$ can be replaced by ${x'}_2^{\mu}$
which is light-like with the initial position $0$.
\item 
Now one observes that the first two steps of the path  $x_i^\mu\rightarrow {x'}_1^\mu\rightarrow
{x'}_2^\mu\rightarrow x_f^\mu$ are on a straight light-curve.
This path is already considered in the construction of the $K^{(1)}_{O}$ TPCF and thus
due to the 
NOC condition arising from Weyl invariance {\bf c)}, it is not to be considered again. This
happens for all the paths one attempts to add for the $K^{(2)}_{O}$ construction.
\item 
The stepwise construction outlined above for $K^{(2)}_{O}$ can by iteration 
be generalized to an arbitrary number of time-like othochronous steps:
$K^{(n)}_{O}$ brings nothing new with respect to $K^{(1)}_{O}$ and thus,
by virtue of the NOC condition $K^{(n)}_{O}$  has to be excluded.  
\end{itemize}
One concludes that the full orthochronous TPCF is
given by the TPCF with one intermediate integration
\bea\label{KO}
K^{}_{O}&=&K^{(1)}_{O}=
{\mathcal{N}} \frac{m \pi}{2} \left(J_0(m |x_f - x_i|)\right.\nonumber\\
&&\left.- i \,H_0(m |x_f - x_i|)\right).
\eea
There is one implicit assumption in this construction in $1+1$ dimensions, which is shown in the subfigure of \ref{fig:twostep1}.
If the second integration point $x_2^\mu$ happens to be below the straight line
connecting ${x'}_1^\mu$ and $x_f^\mu$, the continuous transformation of the point
 ${x}_2^\mu$
to a light cone of the point  ${x'}_1^\mu$ brings it to the point  ${x'}_2^\mu$.
However, the light-lines  ${x'}_1^\mu \rightarrow {x'}_2^\mu $ and
${x}_i^\mu \rightarrow {x'}_1^\mu $ are going into 
 opposite spatial directions.
In order to get the points aligned one needs to spatially flip from ${x'}_2^\mu \Rightarrow {x''}_2^\mu $,
as indicated on the right-hand side of figure \ref{fig:twostep1}.
The assumption one has to make here 
is that this discrete flip forms part of the hidden (local) symmetry
of equivalent paths {\bf{b)}}. This is an isolated problem of $1+1$ dimensions, since in
spatial dimension higher than one {\bf{b)}} contains also spatial rotations, which allow
for a continuous transformation from ${x'}_2^\mu$ to ${x''}_2^\mu$.
Thus, from a higher dimensional perspective, the discrete spatial flip can 
be understood as a continuous rotation in the higher dimensional theory.
Nevertheless, it is interesting to explore the possibility 
that this flip is excluded from the symmetry {\bf{b)}} in $1+1$ dimensions.
This will be done below.

\subsection{Space-like, time-like, orthochronous TPCFs, and the Feynman propagator}
\label{subsec:SLTL}

\begin{figure}[hbt]
 \begin{minipage}{0.45\textwidth}
\centerline{\includegraphics[width=\textwidth]{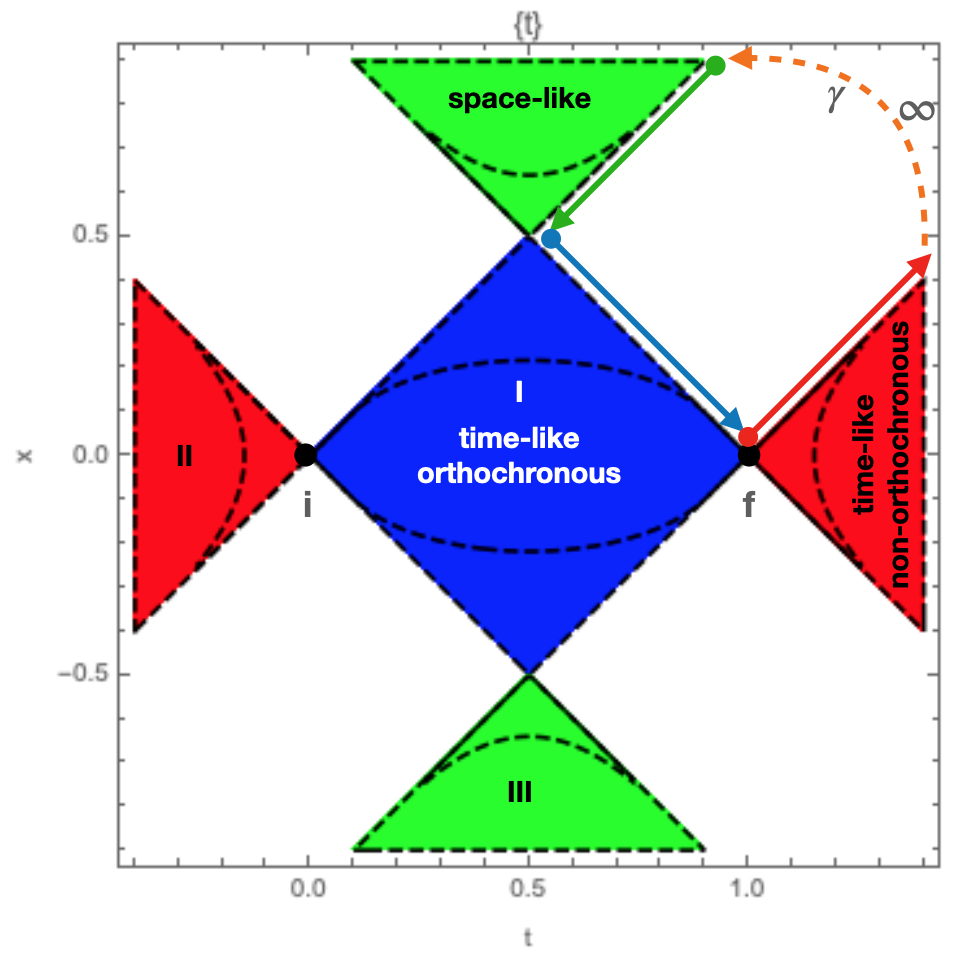}}
\end{minipage}
\ \
\hfill  \begin{minipage}{0.45\textwidth}
\centerline{\includegraphics[width=\textwidth]{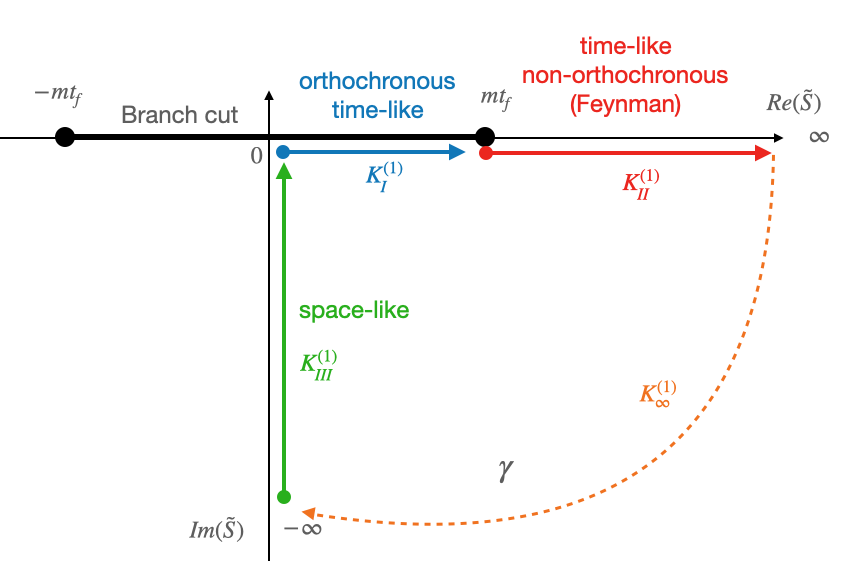}}
\end{minipage}
\ \
\hfill 
  \caption{\label{fig:SecReg} 
  Top: Topology of paths with one intermediate step. Blue stands for time-like orthochronous, red
  for time-like non-orthochronous, and green for space-like. The arrows indicate the integration contour shown in the right panel.\\
  Bottom: Sections and integration contour in the complex $\tilde S$ plane.
  }
\end{figure}

The orthochronous TPCF $K_{O}$ (\ref{res2}) is not the familiar 
position space Feynman propagator of a scalar field in $1+1$ dimensions
\bea\label{KFeyn}
K_F(0,t_f)&=&{\mathcal{N}}\int_1^\infty dy \frac{m}{\sqrt{y^2-1}}e^{-i t_f m y}\nonumber\\
&=&-{\mathcal{N}} \frac{i \pi}{2}H_0^{(2)}(|t_f| m),
\eea
where $H_0^{(2)}$ is the Hankel function. Note that (\ref{KFeyn}) corresponds to the
well known form of the TPCF 
in momentum space 
\be
K_F(k)=\frac{-i }{k^2-m^2 + i \epsilon}.
\ee
To understand the physical meaning of this difference, and how the TPCFs $K_O$
and $K_F$
are connected it is helpful 
to consider the additional space-time regions $V$ for the intermediate steps 
of the TPCF (\ref{K1}).
This is done on the left-hand side of figure \ref{fig:SecReg}.
One can distinguish between three different pure cases
\begin{itemize}
\item[I] The virtual paths are time-like-orthochronous, which means that they are in the future light cone of the initial position and the past light cone of the final position. These paths are indicated
by the blue region of figure \ref{fig:SecReg}.
They lead
to the TPCF $K_I^{(1)}=K_{O}^{(1)}$ given in $\ref{res2}$.
\item[II] The virtual paths are time-like but not orthochronous, which means that the ``intermediate'' step lies either in the future light cone of both $0$ and $t_f$, or it lies in the past light cone of both $0$ and $t_f$. These paths are indicated by the red regions in figure \ref{fig:SecReg}.
Repeating the calculation from subsection \ref{subsecCalc} one finds that 
the integral over this type of paths gives the Feynman propagator of a scalar field 
\be\label{KII}
K_{II}^{(1)}=K_F
\ee
 given in (\ref{KFeyn}) which
  can be seen by the change of integration variables
 $y=\tilde S/(t_f m)$.
\item[III] The third possibility is that all virtual paths are space-like, which 
is indicated by the green regions in figure \ref{fig:SecReg}.
These paths have imaginary action and by using the methods of
subsection \ref{subsecCalc} one finds the TPCF for space-like virtual paths
\bea\label{KIII}
K_S^{(1)}&\equiv& K_{III}^{(1)}=-{\mathcal{N}} \int_0^\infty dy \frac{e^{-t_f m y}}{\sqrt{y^2+1}}\nonumber\\
&=&{\mathcal{N}}\frac{\pi}{2} (Y_0(t_f m)-H_0(t_f m)),
\eea
where $Y_0$ is the Bessel function of the second kind and $H_0$ is the Struve function.
\end{itemize}
There are further regions where the action contains both a real and an imaginary
part. These regions, which are left blank in the left subfigure 
of \ref{fig:SecReg}, would correspond to paths which cross their light-cones along the way.
As shown in appendix \ref{AppB}, such mixed paths can not be
tackled with methods proposed in this paper.
They will not be considered in the construction of the TPCF.
This means that all paths are ``free to do what they want'', as long as
they stay within their respective causal pattern. Note that one can perform
a very similar path integral construction for paths which mix different causal regions,
if one considers in the action (\ref{S1}) absolute values instead of square root contributions for each step.
This will be left to a future study.

For the regions considered here, one realizes that the three TPCFs $K_{I}^{(1)},\, K_{II}^{(1)},\, K_{III}^{(1)}$
all arise from integrating the function
\be\label{fx}
f(y,t_f m)= \frac{1}{\sqrt{y^2-1}}e^{-iy t_f m},
\ee
along different sections of the complex plane. The integration
variable is given from the action by $y=\tilde S/(m t_f)$. Due to the square root in the denominator, this function
possesses a branch cut along the real axis, as depicted in Fig.~\ref{fig:SecReg}. 
Thus, as shown on the right-hand
side of figure  \ref{fig:SecReg}
the integrals over  (\ref{fx})  
can be plugged together to form a closed contour $\gamma=\gamma(y)$.
Since this contour does neither cross the branch cut, nor it does contain zeros or poles
one can apply the residue theorem 
\be\label{residue}
\oint_\gamma dy \, f(y,t_f m)= K_{I}^{(1)}+K_{II}^{(1)}+K_{\infty}^{(1)}+K_{III}^{(1)}=0.
\ee
Since, due to the exponential factor, the angular integral at infinity vanishes $K_{\infty}^{(1)}=0$ one finds
that the Feynman propagator can either be obtained
by an integration over all time-like non-orthochronous paths (\ref{KII}), or
by an integration over all time-like orthochronous paths plus all space-like paths
\be\label{FSOrelation}
K_F(t_f m)=- \left(K_S^{(1)}(t_f m)+ K_O^{(1)}(t_f m)\right).
\ee

The generalization of the results (\ref{KII}) and (\ref{FSOrelation})
for the $n$-step TPCF is straightforward and it follows from the same
arguments already presented in subsection \ref{sec:nprop}, 
if one considers the following adjustments
\begin{itemize}
\item One  connects virtual paths of the same type: e.g. 
orthochronous with orthochronous, space-like with space-like, or non-orthochronous with
non-orthochronous.
\item The discrete symmetry for space-like paths is still a spatial ``flip''
while the discrete symmetry for non-orthochronous time-like paths is a flip from the
past-past to the future-future light cone or vice versa.
\end{itemize}
To summarize, one finds
\bea\label{KIIFull}
K_{I}(t_f m)&=&K_{I}^{(1)}(t_f m)=K_O(t_f m),\\
K_{II}(t_f m)&=&K_{II}^{(1)}(t_f m)=K_F(t_f m)\\
K_{III}(t_f m)&=&K_{III}^{(1)}(t_f m)=K_S(t_f m)
\eea
and
\be\label{FSOrelationFull}
K_F(t_f m)=- \left(K_S(t_f m)+ K_O(t_f m)\right).
\ee
In the light of the above analytic relation, it is interesting to convince oneself geometrically
that if one combines an orthochronous step with a spatial step
some paths can be transformed by the use of {\bf b)} to a non-orthochronous
path appearing in $K_{II}^{(1)}$.

\subsection{Comparison to the TPCFs of the Klein-Gordon field}
\label{subsec:CompKG}

The most common TPCFs for scalar fields are 
the advanced $\Delta_+$, retarded $\Delta_-$, causal $\Delta_C$, Wightman $\Delta_H$, and the Feynman propagator $K_F$.
An important distinction between all these TPCFs is their 
behavior under sign changes of their argument $t_f\leftrightarrow - t_f$.

The relativistic point particle action (\ref{S1}) and its path integrals are blind to such changes $t_f\leftrightarrow - t_f$
and any distinction using this criterion would not be related to
the causal structure of the action discussed in the previous subsection.
Thus, we are working under the assumption of positive $t_f>0$ (or even transformation behavior),
but the resulting TPCFs $K_O$ contains both even and odd functions of $t_f$.

From the above-mentioned TPCFs for the Klein-Gordon equation
only the Feynman propagator has the desired even transformation behavior $K_F(t_f)=K_F(-t_f)$.
Thus, only this TPCF does appear in the comparison to
the results for the relativistic point particle calculated here.
Instead, for example, the causal TPCF $\Delta_C=\Delta_+-\Delta_-$ is
not directly related to the time-like orthochronous TPCF $K_O$.

\subsection{Without spatial flip symmetry and Feynmans' checkerboard}
\label{sec:npropnoflip}

In this  sub-section part of 
the procedure from the previous subsections will be repeated,
but the discrete flips in $1+1$ dimensions will be taken to be as
physically in-equivalent paths and not part of the local symmetry {\bf{b)}}.
The orthochronous TPCF arising from this will be called
$K^{(l)}_{OC}$ as opposed to the $K^{(l)}_O$ TPCF from the previous section.

%
 \begin{figure}
 \begin{minipage}{0.4\textwidth}
\centerline{\includegraphics[width=\textwidth]{step1b.png}}
\end{minipage}
\ \
\hfill  \begin{minipage}{0.4\textwidth}
\centerline{\includegraphics[width=\textwidth]{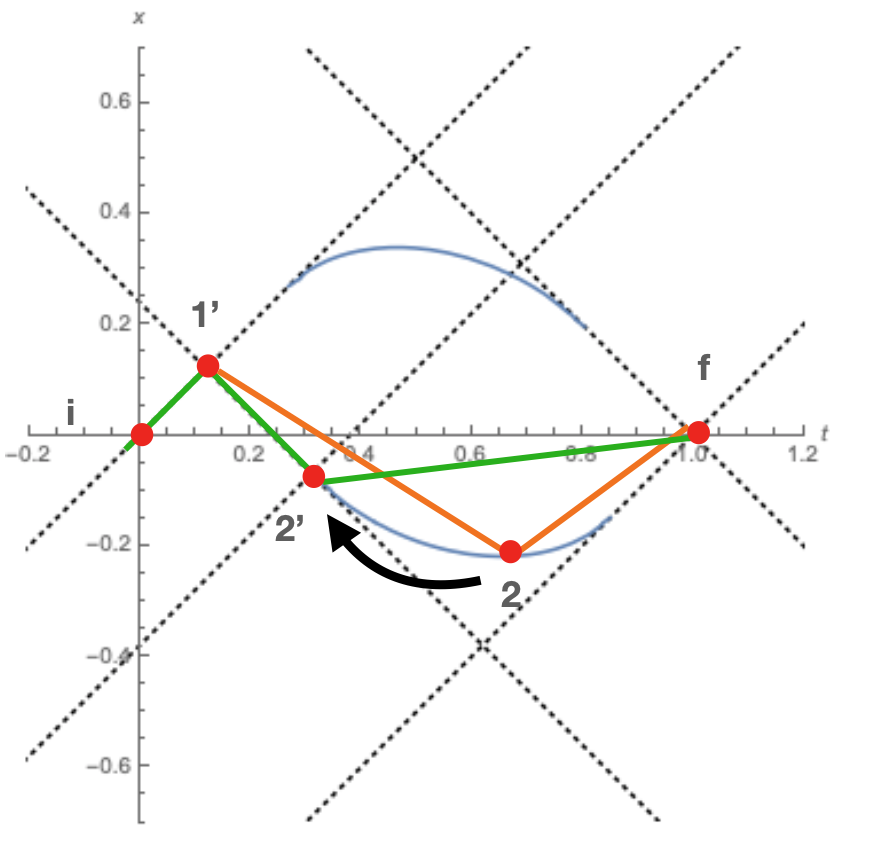}}
\end{minipage}
\ \
\hfill 
\begin{minipage}{0.4\textwidth}
\centerline{\includegraphics[width=\textwidth]{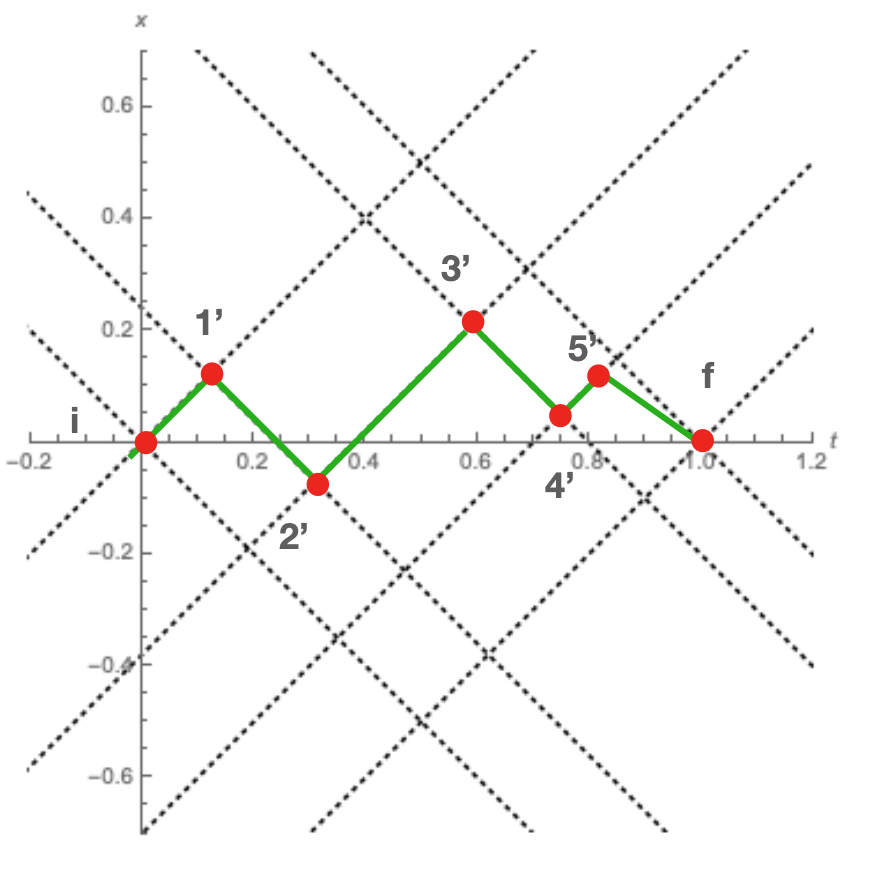}}
\end{minipage}
\ \
\hfill 
  \caption{\label{fig:twostep2}
The topological construction of the paths leading to the Orthochronous TPCF is displayed. 
}
\end{figure}
%

The construction goes as follows
\begin{itemize}
\item For the TPCF $K^{(1)}_{OC}$ one chooses the first point
${x}_1^\mu$ and uses the symmetry {\bf b)} to shift
to a physically equivalent point ${x}_1^\mu \Rightarrow {x'}_1^\mu$.
Already here one can distinguish between $K^{(1)}_O$ going up
and  $K^{(1)}_O$ going down, which have to be summed.
This is shown in the left panel of figure \ref{fig:twostep2}.
\item For the TPCF  $K^{(2)}_{OC}$ one chooses the second point
${x}_2^\mu$, while taking ${x'}_1^\mu$ as given and uses the symmetry {\bf b)} to shift
to a physically equivalent point ${x}_2^\mu \Rightarrow {x'}_2^\mu$,
which is light-like with ${x'}_1^\mu$.
If this point is the straight continuation of the light-line 
${x}_i^\mu \rightarrow {x'}_1^\mu$, the discussion from the previous section
applies and this contribution is discarded. 
If, to the contrary, light-lines ${x}_i^\mu \rightarrow {x'}_1^\mu$ and 
${x'}_1^\mu \rightarrow {x'}_2^\mu$ are going in opposite spatial directions,
as shown in the middle panel of figure \ref{fig:twostep2},
this path is a new contribution which forms part of $K^{(2)}_{OC}$.
\item For the TPCF $K^{(l)}_{OC}$, the construction continues analogously.
Countable new paths only occur when there is a kink between two subsequent light-lines.
For $K^{(5)}_{OC}$ this is shown in the right panel of figure \ref{fig:twostep2}.
\end{itemize}

One realizes that for each step from $K^{(l)}_{OC}$ to $K^{(l+1)}_{OC}$, the 
available maximal absolute value of the action $\tilde S_{(l+1)}$ reduces, such that for
$\lim_{l\rightarrow \infty} K^{(l)}_{OC}$ the entire path is light-like.

The total TPCF is now
\be\label{KOC1}
K_{OC}(x_i^\mu,x_f^\mu)=\sum_{l=1}^\infty K_{OC}^{(l)}(x_i^\mu,x_f^\mu),
\ee
which is harder to work out than the TPCF (\ref{KO}).
First, as in sub-section \ref{subsecCalc} 
one can write out the integral expression for each of the TPCFs
$K_{OC}^{(l)}$ in terms of the  absolute value of the corresponding action
\bea
&&K_{OC}(x_i^\mu,x_f^\mu)=\lim_{N\rightarrow \infty} \sum_{l=1}^N
\omega^l\nonumber\\
&&\times\prod_{j=1}^{l} \left( \int_0^{\tilde S_{j-1}} dS_j \frac{1}{\sqrt{\tilde S_{j-1}^2-\tilde S_j^2}}\right)
e^{-i (\tilde S_l)},
\eea
where $\tilde S_0=m \sqrt{(x_i-x_f)^\mu (x_i-x_f)_\mu}$
and $\omega$ is the measure weight associated to each change in direction,
where a discrete realization of {\bf b)} was avoided.
This is a sum over all light-like paths with $l$ turns,
where only one of the steps is time-like. In the right
panel of figure \ref{fig:twostep2} this step was chosen as the last one
$x_5'\rightarrow x_f$. By approaching the construction from the
left and from the right the non-light-like step could have been any
pair ${x'}_{j-1}^\mu\rightarrow {x'}_j^\mu$.
In any case,  for large $l\rightarrow \infty$
the typical value for the remaining action $\sim \sqrt{({x'}_{j-1}- {x'}_j)^\mu ({x'}_{j-1}- {x'}_j)_\mu}$ 
would approach zero $\tilde S_l\rightarrow 0$.

Thus, for a large number of steps the TPCF $K_{OC}^{(l)}$ contains a discrete sum
over all light-like paths with the additional measure weight $\omega^l$.
\be
K_{OC}(x_i^\mu,x_f^\mu)=\lim_{N\rightarrow \infty} \sum_{l=1}^N
\omega^l \Phi(l,N),
\ee
where $\Phi(l,N)$ is the number of paths with $l$ turns for a given discretization $N$.

A beautiful realization of this concept is given in terms of the Feynman checkerboard,
where the weight is chosen to be \cite{Jacobson:84}
\be
\omega = i \frac{m c^2}{\hbar}\frac{(t_f-t_i)}{N}.
\ee
Conceptually, this choice for the weight factor arises as a consequence of Heisenberg's uncertainty principle, $\Delta x \Delta p \ge \hbar$.
For a massive relativistic particle, the minimum uncertainty in momentum is $\Delta p \sim m c$, and therefore the leaps that
the particle experiences should satisfy $\Delta x \ge \hbar/(mc)$, i.e. they cannot be smaller than the Compton wavelength $\bar{\lambda}_C$. For a finite
path consisting on a large number of such "leaps" with $\Delta x \simeq c \Delta t$, the ratio $\omega = i\Delta x/\bar{\lambda}_C = i m c^2 \Delta t/\hbar$ thus provides the apropiate
weight such that the path-integral is dominated by those trajectories where the uncertainty principle inequality is satisfied in the majority of steps.

In the checkerboard $\Phi(l,N)$ is approximated by
a lattice of $N$ light-like steps with $l$ turns of the spatial direction.
Further, a distinction  between the direction (+,-) of the first and last step is made.
With this distinction the TPCF takes the form of a  $2\times 2$
matrix
\be\label{propCheck}
K_{OC}(x_i^\mu,x_f^\mu)=
\lim_{N\rightarrow \infty} \sum_{l=1}^N \omega^l
\begin{pmatrix}
\Phi_{++}(l,N)&\Phi_{+-}(l,N)\\
\Phi_{-+}(l,N)&\Phi_{--}(l,N)
\end{pmatrix},
\ee
where for example $\Phi_{++}(l,N)$ stands for the number of paths that
leave moving right and arrive moving right.
Interestingly, (\ref{propCheck}) turns out to be identical
to the Dirac propagator in position space  $1+1$ dimensions. As discussed in detail in \cite{Jacobson:84}, the problem of calculating the coefficients $\Phi_{\alpha\beta}(l,N)$
can be cast in the form of an Ising model, defining the instantaneous direction of the j-th step by a two-valued spin $\sigma_j = \pm$, such that the net number of steps in the positive
$x-direction$ is
\begin{eqnarray}
M = \sum_{j = 1}^{N}\sigma_j,
\end{eqnarray}
whereas the total number of flips $l$ corresponds in this language to
\begin{eqnarray}
l = \frac{1}{2}\sum_{j = 1}^{N-1}\left( 1 - \sigma_j\sigma_{j+1}  \right) = \frac{N-1}{2} - \frac{1}{2}\sum_{i = 1}^{N-1}\sigma_j\sigma_{j+1}.
\end{eqnarray}
Therefore, after introducing an integral representation for the Kronecker delta,
\begin{eqnarray}
\delta_{M,\sum_j \sigma_j} = \int_{-\pi}^{\pi}\frac{d\theta}{2\pi} e^{i\theta\left( \sum_{j=1}^{N}\sigma_j - M  \right)}
\end{eqnarray}
the elements of the TPCF Eq.~(\ref{propCheck}) correspond to the sum (for $\alpha = \pm$ and $\beta=\pm$ not contracted, but fixed and in correspondence with $\Phi_{\alpha\beta}$)
\begin{eqnarray}
\sum_{l=1}^{N}\Phi_{\alpha\beta}(l,N)\omega^l &=& \sum_{\sigma_2 = \pm}\ldots\sum_{\sigma_{N-1}=\pm} \omega^l\\
&=& \int_{-\pi}^{\pi}\frac{d\theta}{2\pi} e^{i M \theta}e^{-\frac{i}{2}\theta(\alpha + \beta)} \left[ \hat{T}^{N-1} \right]_{\alpha\beta},\nonumber
\end{eqnarray}
where by analogy with the Ising model, the transfer matrix is defined by
\begin{eqnarray}
\hat{T}_{\sigma,\sigma'} = \exp\left(\nu \sigma\sigma' + i\frac{\theta}{2}\left( \sigma + \sigma' \right) - \nu  \right),
\end{eqnarray}
and $\nu = -\frac{1}{2}\log\left( \omega \right)$.
Applying a similarity transformation that diagonalizes the transfer matrix, a saddle point calculation is performed to determine the stationary value of the parameter $\sin\theta = i\mu\omega v/\sqrt{1 - v^{2}}$,
with $\mu = \pm 1$ and $v = \left( x_f - x_i \right)/(-i N \omega)$ the linear "speed" of the particle (in natural units such that $c = \hbar = 1$). Therefore, the corresponding expression obtained
for the propagator from this procedure is the famous checkerboard result~\cite{Jacobson:84}
\begin{widetext}
\begin{eqnarray}
\sum_{l = 1}^{N}\Phi_{\alpha\beta}(l,N)\omega^l &=& \omega \sum_{\mu=\pm}\int\frac{dp}{2\pi}\frac{1}{2}\left[ 1 + \mu\frac{m\sigma_x - p\sigma_z}{\sqrt{p^2 + m^2}} \right]_{\alpha\beta}
e^{ip(x_f - x_i)}e^{i\mu \sqrt{p^2 + m^2}}=\frac{\omega}{2\pi} \int dp e^{i p (x_f - x_i)} \left[e^{ i \left( m\sigma_x - p\sigma_z \right)} \right]_{\alpha\beta}
\end{eqnarray}
\end{widetext}

\subsection{Higher dimensional generalization}
\label{sec:high_dim}

The higher dimensional generalization of calculation of the (1+1) dimensional TPCF
can be done straight forwardly. To vary the construction, the PMS will be applied.
In (1+d) dimensions the TPCF (\ref{K1}) reads
\be\label{K1ddim}
K^{(1)}={\mathcal{N}}\cdot  \int dt_1  \int_V d^dx_1 \Xi_1 e^{i S_1},
\ee
First, one can write the $d$ spatial integrals in spherical coordinates
\be\label{K1ddim2}
K^{(1)}={\mathcal{N}}\cdot  \int dt_1  \int_V d|x_1| d\Omega_{d-1} |x_1|^{d-1} \Xi_1 e^{i S_1}.
\ee
The integrand is independent of the angular coordinates and thus,
the integral over $d\Omega_{d-1}$ gives $2 \pi^{(d-1)/2}/\Gamma((d-1)/2)$, which is
the surface of the unit sphere in $d-1$ dimensions.
This constant factor can be absorbed into the normalization constant.
Now one proceeds with a change in the integration variables from $|x_1|\rightarrow S$
given in (\ref{coordtrans}),
keeping the $t_1$ integration untouched.
The integral now reads
\be\label{K1ddim2}
K^{(1)}={\mathcal{N}}\cdot   \int_{-t_f m}^{0} dS \int_{t_f/2-S^2/(2 m^2 t_f)}^{t_f/2+S^2/(2 m^2 t_f)} dt_1  J \cdot |x_1|^{d-1} \Xi_1 e^{i S}.
\ee
After a change of variables $S = -\tilde{S}$, we obtain
\be\label{K1dddim2}
K^{(1)}={\mathcal{N}}\cdot   \int_{0}^{t_f m} d\tilde{S} \int_{t_f/2-\tilde{S}^2/(2 m^2 t_f)}^{t_f/2+\tilde{S}^2/(2 m^2 t_f)} dt_1  J \cdot |x_1|^{d-1} \Xi_1 e^{-i \tilde{S}}.
\ee
The factor $J \cdot |x_1|^{d-1}$ is a function of $\tilde{S}$, where $|x_1|$ is given by (\ref{coordtrans})
and
\be
J =\frac{m^4 t_f^2 (t_f-2 t_1)^2-\tilde{S}^4}{2 m \tilde{S}^2 \sqrt{(m^2 t_f^2-\tilde{S}^2)(\tilde{S}^2-m^2(t_f-2 t_1)^2)}}.
\ee
Now one has to get rid of the $t_1$ dependence.  
Instead of imposing a $t_1$ dependent Fujikawa factor $\Xi_1=\Xi_1(t_1)$, we will now choose
an optimal $t_1$ such that
\be
\frac{d}{dt_1} J\cdot |x_1|^{d-1}|_{t_1=t_{1,opt}}=0
\ee 
The solutions to this condition are
\be
t_{1,opt}= \left \{
\begin{array}{c}
\frac{t_f}{2}\\
\frac{m t_f \pm \tilde{S}}{2 m}\\
\frac{t_f}{2}\pm \frac{m^4 \tilde{S}^2 t_f^2 (2 m^2 t_f^2+\tilde{S}^2d)d}{2 m^4 t_f^2d}
\end{array}
\right. .
\ee
One realizes that the only solution which lies in the allowed integration
range of $dt_1$ is 
\be
t_{1,opt}=\frac{t_f}{2}.
\ee
With this, the TPCF reads, up to a normalization constant
\be\label{K1ddim2}
K^{(1)}={\mathcal{N}}\cdot   \int_0^{t_f m} d\tilde{S}  \left( m^2 t_f^2-\tilde{S}^2 \right)^{(d-2)/2} \tilde{S} \Xi_1 e^{-i \tilde{S}}.
\ee
According to the condition $\alpha)$, this has to be finite for $\tilde{S} \rightarrow 0$ and with the right dimensions,
which gives
\be
\Xi_1=\frac{1}{\tilde{S} (t_f m)^{d-2}}
\ee
and thus
\be\label{K1ddim3}
K^{(1)}={\mathcal{N}}\cdot   \int_0^{t_f m} d\tilde{S}  \left(1-\frac{\tilde{S}^2}{m^2 t_f^2} \right)^{(d-2)/2}  e^{-i \tilde{S}},
\ee
which is the TPCF in $1+d$ dimensions.
In order to bring this to a more familiar form one can perform a number of operations. The coordinate
transformation
\be
\tilde{S}=y m t_f
\ee
gives
\be\label{K1ddim4}
K^{(1)}={\mathcal{N}}\cdot   \int_0^{1} dy  \sqrt{y^2 -1}^{(d-2)}  e^{-i m t_f y},
\ee
where a factor of $i^{d-2} $ was absorbed in the normalization.
Now one defines
\be
y = \frac{\sqrt{\vec k^2+m^2}}{m},
\ee
which leads to
\be\label{K1ddim4}
K^{(1)}={\mathcal{N}}\cdot   \int_0^{im} dk  \frac{k^{d-1} e^{-i t_f \sqrt{\vec k^2+m^2}}}{\sqrt{\vec k^2+m^2}},
\ee
The $dk$ integral can be written in terms of a $d$ dimensional integral
\be\label{K1ddim5}
K^{(1)}={\mathcal{N}}\cdot   \int_0^{im} d^{d}k  \frac{ e^{-i |t_f| \sqrt{\vec k^2+m^2}}}{\sqrt{\vec k^2+m^2}},
\ee
where the assumption of $t_f>0$ was made explicit by writing $|t_f|$
and where the upper limit is for the radial component of $dk$.
This absolute value can then be written in terms of a $\theta$ function prescription,
which in turn can be written as an integral over $k_0$ leaving 
\be
\label{K1ddim6}
K^{(1)}={\mathcal{N}}\cdot  \int_{-\infty}^{\infty} dk_0  \int_0^{im} d^{d}k  \frac{ e^{-i k_0 t_f}}{k^2-m^2+i\epsilon}.
\ee
This is the familiar momentum space representation of the TPCF, but with a different contour of integration,
as explained before.
The analytical expression for this integral, up to action-independent constants absorbed in the normalization coefficient, is
\begin{eqnarray}
K^{(1)}_{O}(0,t_f)&=&{\mathcal{N}}\cdot \left( m t_f\right)^{\frac{1-d}{2}}\left[ J_{\frac{d-1}{2}}\left( m t_f \right)\right.\nonumber\\
&&\left.- i  H_{\frac{d-1}{2}}\left( m t_f \right)  \right], 
\end{eqnarray}
where $J_{\frac{d-1}{2}}$ is the Bessel function of the first kind, and $H_{\frac{d-1}{2}}$ is the Struve function, and we used again the "orthochronous" prescription, generalizing Eq.~(\ref{res2}).
By the same arguments presented in the previous section, the generalization of the Feynman propagator to $1+d$ dimensions is
\begin{eqnarray}
K^{(1)}_F(0,t_f) &=& {\mathcal{N}}\cdot   \int_1^{\infty} dy  \sqrt{y^2 -1}^{(d-2)}  e^{-i m t_f y}\nonumber\\
&=& {\mathcal{N}}\left(m t_f  \right)^{\frac{1-d}{2}} H_{\frac{d-1}{2}}^{(1)}\left( m t_f \right),
\label{KFddim4}
\end{eqnarray}
where $H_{\frac{d-1}{2}}^{(1)}$  is the Hankel function of order $(d-1)/2$, and all action-independent constants have been absorbed into the overall normalization.
Finally, the generalized form of the TPCF for space virtual paths is given by (see Appendix for details)
\begin{eqnarray}\label{KSddim4}
K^{(1)}_S(0,t_f) &=& -{\mathcal{N}}\cdot   \int_0^{\infty} dy  \sqrt{y^2 +1}^{(d-2)}  e^{- m t_f y}\\
&=&  {\mathcal{N}}\cdot \left( m t_f \right)^{\frac{1-d}{2}}\left[
Y_{\frac{d-1}{2}}(m t_f) -  H_{\frac{d-1}{2}}(m t_f)
\right],\nonumber
\end{eqnarray}
where $Y_{\frac{d-1}{2}}$ is the Bessel function of the second kind.
The proof for $N$ steps is completely analogous to the lower dimensional case, so it will not be repeated.

\subsection{Higher dimensional generalization of the checkerboard?}
\label{sec:high_dimCB}

The higher dimensional generalization shown in the previous subsection
works out so nicely, that one is tempted to expect a similar result for the 
Feynman checkerboard, discussed in subsection \ref {subsec:SLTL}.
Certainly, the checkerboard approach has been explored in dimensions higher than $(1+1)$~\cite{Mckeon:1993,Smith:1995kd,Kull:1999vz}.
Nevertheless, the elegance and interpretation in terms of a simple summation over light-like paths is lost in these attempts.
This unexpected fact can be easily understood in terms of the symmetry construction presented in this paper.
As shown above, the checkerboard construction in $(1+1)$ dimensions arises naturally from counting
discrete parity bounces as independent paths contributing to the TPCF. This was justified,
since two such paths are not connected by a continuous symmetry transformation and thus,
there is no overcounting in the sense of the conditions ({\bf a})-({\bf c}). 
However, in dimensions higher than $(1+1)$ any of these discrete bounces
(see eg. figure \ref{fig:twostep2} )
can be generated from a continuous rotation of the spatial directions. 
Such continuous transformations are
overcountings in the sense of the conditions ({\bf a})-({\bf c}) and they
have to be factored out. This is the reason why insisting on a checkerboard construction with more than one
spatial dimension is unnatural.

\section{Conclusion}

In this paper, we studied the path-integral construction of the TPCFs of the free relativistic point particle. After analyzing the global and local symmetries of the action, we identified
a local ``hidden'' symmetry that is usually disregarded in the literature. 
This symmetry corresponds to the invariance of the modulus of the 4-velocity with respect to Lorentz boosts and rotations.
By taking explicit care of this symmetry, and removing the associated redundant phase-space volume that leads to the overcounting of trajectories in the path-integral by means of a Fujikawa prescription, we were able to
obtain the correct results for the TPCFs that recover the Chapman-Kolmogorov property. Furthermore, we formulated the path-integral explicitly in Minkowski space, by carefully taking into account
the non-simply connected structure of the Lorentz group. In a detailed topological analysis for the case of $D = 1 + 1$ dimensions, we explicitly 
constructed different versions of the relativistic TPCF, for the different causal characteristics of the virtual paths.
As a new result, we obtained a causal-orthochonous- and a space-like-TPCF and we also recovered the Feynman propagator.
It was further shown that these three TPCFs are related due to a closed contour integral in the (complex-valued) Minkowski space.
By means of this identity, the Feynman propagator could either be understood as the result of a PI over time-like but non-orthochronous paths,
or it could be understood as PI over space-like and time-like-orthochronous paths.
 The TPCFs in subsections 
\ref{subsecCalc} and \ref{sec:nprop} were straight forwardly generalized to higher dimensions $D = d + 1$.

Finally, based on the methods presented before, it has been shown how the Feynman checkerboard construction in $1+1$ dimensions 
arises naturally within this framework
and it has also been explained why a natural generalization of the checkerboard approach to higher dimensions does not work.

\section*{Acknowlegements}

E.~M. was supported by  Fondecyt Regular No 1190361, and by PIA
Anillo ACT192023.

\appendix

\section{}\label{AppA}

In this appendix, we show that the different Bessel and related functions (such as Struve and Hankel)
whose linear combinations give rise to the TPCFs obtained via path-integration in this work,
when their argument depends on the time-like interval $|x| =\sqrt{ c^2 t^2 - \mathbf{x}^2}$, are themselves solutions of the
Klein-Gordon equation for $|x|>0$.

In mathematical terms, we shall prove the following simple theorem,

{ \it{Theorem}: If $y_{\alpha}(z)$ satisfies the Bessel equation
\begin{eqnarray}
z^2\frac{d^2y_{\alpha}}{dz^2} + z\frac{dy_{\alpha}}{dz} + \left( z^2 - \alpha^2 \right) y_{\alpha}(z) = 0,
\label{eq_A1}
\end{eqnarray}
then for $\alpha = 0$ the function $y_{0}(z = m|x|)$ satisfies the Klein-Gordon equation}
\begin{eqnarray}
\left(\square_x + m^2\right) y_{0}(m|x|) = 0.
\label{eq_A2}
\end{eqnarray}
For simplicity, the proof will be given in spatial dimension $d = 1$, where we have $|x|^2 = c^2 t^2 - x^2$. Therefore, the following
simple relations follow after application of the chain rule for partial derivatives
\begin{eqnarray}
\frac{1}{c}\frac{\partial |x|}{\partial t} &=& \frac{c t}{|x|}\nonumber\\
\frac{1}{c^2}\frac{\partial^2 |x|}{\partial t^2} &=& \frac{1}{|x|}\left( 1 - \frac{c^2 t^2}{|x|^2} \right) = -\frac{x^2}{|x|^3},
\label{eq_A3}
\end{eqnarray}
and similarly
\begin{eqnarray}
\frac{\partial|x|}{\partial x} &=& -\frac{x}{|x|}\nonumber\\
\frac{\partial^2|x|}{\partial x^2} &=& -\frac{1}{|x|}\left( 1 + \frac{x^2}{|x|^2} \right) = -\frac{c^2 t^2}{|x|^3}.
\label{eq_A4}
\end{eqnarray}
Let us now consider the action of the d'Alembert operator over a function $y_0(z=m|x|)$ that, by hypothesis,
satisfies Eq.~(\ref{eq_A1}). Therefore, after some straightforward algebra, we obtain
\begin{eqnarray}
\square_x y_{0}(m|x|) &=& \frac{1}{c^2}\frac{\partial^2}{\partial t^2}y_{0}(z) - \frac{\partial^2}{\partial x^2}y_{0}(z)\nonumber\\
&=& \frac{m}{c^2}\frac{\partial}{\partial t}\left( \frac{\partial |x|}{\partial t} \frac{d y_0(z)}{dz} \right)- m \frac{\partial}{\partial x}\left( \frac{\partial|x|}{\partial x} \frac{d y_0(z)}{dz} \right)\nonumber\\
&=& m\left( \frac{1}{c^2}\frac{\partial^2|x|}{\partial t^2} - \frac{\partial^2|x|}{\partial x^2} \right)\frac{dy_0}{dz}\nonumber\\
&+& m^2\left( \frac{1}{c^2}\left( \frac{\partial|x|}{\partial t} \right)^2 - \left( \frac{\partial|x|}{\partial x} \right)^2 \right)\frac{d^2y_0}{dz^2}
\label{eq_A5}
\end{eqnarray}
Substituting Eq.~(\ref{eq_A3}) and Eq.~(\ref{eq_A4}) into Eq.~(\ref{eq_A5}), and simplifying, we obtain
\begin{eqnarray}
\square_x y_{0}(m|x|) &=& m^2 \frac{d^2y_0}{dz^2} + \frac{m}{|x|}  \frac{dy_0}{dz}\nonumber\\
&=& m^2\left[ \frac{d^2y_0}{dz^2} + \frac{1}{m |x|}  \frac{dy_0}{dz} \right]\nonumber\\
&=& m^2\left[ \frac{d^2y_0}{dz^2} + \frac{1}{z}  \frac{dy_0}{dz} \right]\nonumber\\
&=& -m^2 y_0(z),
\label{eq_A6}
\end{eqnarray}
where in the last line we applied that, by hypothesis, the function $y_0(z)$ is a solution to the Bessel differential Eq.~(\ref{eq_A1}) for $\alpha = 0$. This proves
the Theorem, i.e. that $y_0(m|x|)$ is indeed a solution of the Klein-Gordon Eq.~(\ref{eq_A2}) if it satisfies Eq.~(\ref{eq_A1}).

Since all the TPCFs obtained in this work after the path-integral procedures described in the text lead to linear combinations of Bessel and related functions (with $\alpha = 0$),
and since each of those functions satisfy the conditions of the theorem, we conclude that each of the TPCFs constitute different solutions of the Klein-Gordon equation
\begin{eqnarray}
\left( \square_x + m^2 \right) K(m |x|) = 0.
\label{eq_A7}
\end{eqnarray}
\vspace{-0.2cm}

\section{}\label{AppB}

Intermediate points $(t_1,x_1)$ of the two step construction that are equivalent in terms of the hidden symmetry~({\bf{b}})
are shown as colored lines in figure~\ref{LT0} and as dashed lines
in the upper part of figure~\ref{fig:SecReg}. Along these lines
the action (\ref{S1}) is constant and contains two contributions
\bea
S_1&= &S_{01}+ S_{1f}\\ \nonumber
&=&\sqrt{t_1^2-x_1^2}+ \sqrt{((t_1-t_f)-x_1)((t_1-t_f)+x_1)},
\eea
where  we chose without loss of generality~$t_i=x_i=x_f=0$.
If both contributions $S_{01}$ and $S_{1f}$ are real, the path is time-like (non)-orthochronous 
leading to $K_{I}^{(1)}$ ($K_{II}^{(1)}$).
If both contributions are imaginary, the path is space-like leading to $K_{III}^{(1)}$.
In these sections (I, II, III), the direction of the tangential vector $(\xi_t, \xi_x)$,  generating  the hidden-gauge-equivalent lines, is obtained from the condition of constant action
\be \label{ConstSCond}
\xi_t (\partial_{t_1} S_1) + \xi_x (\partial_{x_1} S_1)=0.
\ee
One finds
\be\label{ConstSsol}
\frac{\xi_t}{\xi_x}=\frac{x_1S_1}{t_1 S_{1f}+ S_{01} (t_1-f_f)}.
\ee
There is also the possibility that one contribution (e.g. $S_{01}$) to a path is
real and the other contribution is imaginary (e.g. $S_{1f}$).
These intermediate points are the blank regions in figure~\ref{fig:SecReg}.
For a hidden symmetry of the type~({\bf{b}}) to exist, one needs
a constant real and a constant imaginary part of the total action.
Thus, there are two conditions of the type (\ref{ConstSCond}),
one for the real part and one for the imaginary part
\bea \label{ConstSCond2}
\xi_t (\partial_{t_1} S_{01}) + \xi_x (\partial_{x_1} S_{01})&=&0, \\ \nonumber
\xi_t (\partial_{t_1} S_{1f}) + \xi_x (\partial_{x_1} S_{1f})&=&0.
\eea
This leads to the two relations
\bea
\frac{\xi_t}{\xi_x}&=&\frac{x_1}{t_1}\\
\frac{\xi_t}{\xi_x}&=&\frac{x_1}{t_1-t_f}.
\eea
These conditions, in contrast to (\ref{ConstSsol}), can not be fulfilled simultaneously, meaning
that there are no lines of constant action in regions with mixed time-like and space-like
contributions. The Fujikawa techniques proposed in this paper can not be used for such paths. One possibility to avoid this restriction would be to work with the absolute
value of the action. This is not considered in this work.

\section{}\label{AppC}
Here we present the mathematical details leading to Eq.~(\ref{KSddim4}) in the main text. The integral
\begin{eqnarray}
K^{(1)}_S(0,t_f) &=& -{\mathcal{N}}\cdot   \int_0^{\infty} dy  \sqrt{y^2 +1}^{(d-2)}  e^{- m t_f y}\\
&=&  -{\mathcal{N}}\frac{2^{(d-2)/3}\pi^{3/2}}{\Gamma(1 - d/2)}\cdot \left( m t_f \right)^{\frac{1-d}{2}}\left[
\sec\left(\frac{\pi d}{2} \right)\right.\nonumber\\
&&\left.\cdot J_{\frac{d-1}{2}}(m t_f) - 2\csc(\pi d)J_{\frac{d-1}{2}}(m t_f)
\right.\nonumber\\
&&\left.+ \csc\left( \frac{\pi d}{2}\right) H_{\frac{d-1}{2}}(m t_f)
\right].\nonumber
\label{A1}
\end{eqnarray}
Let us consider the following trigonometric identities
\begin{eqnarray}
\sec\left( \frac{\pi d}{2} \right) &=& \frac{1}{\cos\left( \frac{\pi d}{2} \right)} = -\frac{1}{\sin\left( \frac{\pi(d - 1)}{2} \right)},\nonumber\\
\csc\left(\frac{\pi d}{2} \right) &=& \frac{1}{\sin\left(\frac{\pi d}{2} \right)} = \frac{1}{\cos\left(\frac{\pi (d-1)}{2} \right)},\nonumber\\
\csc(\pi d) &=& \frac{1}{2 \sin\left(\frac{\pi d}{2} \right) \cos\left(\frac{\pi d}{2} \right)}\nonumber\\
&=& -\frac{1}{2\sin\left( \frac{\pi(d - 1)}{2} \right)\cos\left( \frac{\pi(d - 1)}{2} \right)}.
\label{A2}
\end{eqnarray}
Substituting these identities, we have
\begin{widetext}
\begin{eqnarray}
K^{(1)}_S(0,t_f) =  {\mathcal{N}}\frac{2^{(d-2)/3}\pi^{3/2}}{\Gamma(1 - d/2)\cos\left(\frac{\pi (d-1)}{2} \right)}\cdot \left( m t_f \right)^{\frac{1-d}{2}}\left[
\frac{\cos\left(\frac{\pi (d-1)}{2} \right) J_{\frac{d-1}{2}}(m t_f) - J_{\frac{1-d}{2}}(m t_f)}{\sin\left( \frac{\pi(d-1)}{2}\right)}
-H_{\frac{d-1}{2}}(m t_f)
\right].
\label{A3}
\end{eqnarray}
\end{widetext}
Finally, using the Bessel function identity
\begin{eqnarray}
Y_{\alpha}(z) = \frac{J_{\alpha}(z)\cos(\alpha\pi) - J_{-\alpha}(z)}{\sin(\alpha\pi)},
\end{eqnarray}
and absorbing all the action-independent constants in the normalization $\mathcal{N}$, we obtain
\begin{equation}
K^{(1)}_S(0,t_f)=  {\mathcal{N}}\cdot \left( m t_f \right)^{\frac{1-d}{2}}\left[
Y_{\frac{d-1}{2}}(m t_f) -  H_{\frac{d-1}{2}}(m t_f)
\right].
\end{equation}



\end{document}